\documentclass{aa}
\input{epsf}

\newcommand{\bb}{\bibitem[]{bla}}
\newcommand{\zm}{ \relax \ifmmode {\rm M_{\odot}} \else {M$_{\odot}$}\fi}

\newcommand{\ang}{$\rm \AA$}

\newcommand{\degree}{$^{\rm o}$}

\newcommand{\ea}{{et al.}}

\newcommand{\km}{km s$^{-1}$}
\newcommand{\ha}{H$\alpha$}

\def\lesssim{\mathrel{\hbox{\rlap{\hbox{\lower4pt\hbox{$\sim$}}}\hbox{$<$}}}}
\def\gtrsim{\mathrel{\hbox{\rlap{\hbox{\lower4pt\hbox{$\sim$}}}\hbox{$>$}}}}

\def\ion#1#2{#1$\;${\small\rm\@Roman{#2}}\relax}
 
\newcommand{\hzz}{HD~76534}

\newbox\grsign \setbox\grsign=\hbox{$>$} \newdimen\grdimen 
\grdimen=\ht\grsign
\newbox\simlessbox \newbox\simgreatbox
\setbox\simgreatbox=\hbox{\raise.5ex\hbox{$>$}\llap
     {\lower.5ex\hbox{$\sim$}}}\ht1=\grdimen\dp1=0pt
\setbox\simlessbox=\hbox{\raise.5ex\hbox{$<$}\llap
     {\lower.5ex\hbox{$\sim$}}}\ht2=\grdimen\dp2=0pt

\makeatletter 
\renewcommand\@biblabel[1]{}     

\makeatother

\begin{document}

\thesaurus{08.01.2 ; 08.03.1 : 08.05.2 ; 08.09.2: HD 76534 ; 08.13.2 ; 08.22.3}

\title
{
{\it Research Note} \, \, \, \, \, \, \, \, \, 
\, \, \, \, \, \, \, \, \, \, \, \, \, \, \, \, \, \, 
\, \, \, \, \, \, \, \, \, \, \, \, \, \, \, \, \, \, 
\, \, \, \, \, \, \, \, \, \, \, \, \, \, \, \, \, \, 
Time-resolved spectroscopy of the  peculiar \ha\ variable  Be star HD~76534
}

\author{ Ren\'e D. Oudmaijer and Janet E. Drew }
\offprints{Ren\'e D. Oudmaijer (e-mail: r.oudmaijer@ic.ac.uk) }

\institute{
Imperial College of Science, Technology and Medicine,
Blackett Laboratory, Prince Consort Road,\\ London,  SW7 2BZ, U.K.   
}

\date{received,  accepted}

\maketitle

\markboth{R.D. Oudmaijer \& J.E. Drew}{Time resolved spectroscopy of the \ha\ variable Be star HD 76534}

\begin{abstract}

We present time-resolved spectroscopy of the Be star HD 76534, which
was observed to have an \ha \ outburst in 1995, when the line went
from photospheric absorption to emission at a level of more than two
times the continuum within 2.5 hours.  To investigate the short-term
behaviour of the spectrum of \hzz \ we have obtained 30 spectra within
two hours real-time and searched for variations in the
spectrum. Within the levels of statistical significance, no
variability was found. Rather than periodic on short time scales, the
\ha \ behaviour seems to be commonly episodic on longer ($>$ 1 year)
time scales, as an assessment of the existing data on the \ha \ line
and the Hipparcos photometry suggests.  HD 76534 underwent only 1
photometric outburst in the 3 year span that the star was monitored by
the Hipparcos satellite.

\end{abstract}

\begin{keywords}
stars: activity  --
stars: circumstellar matter --
stars: emission-line, Be --
stars: individual: \hzz  --
stars: mass loss --
stars: variables: other
\end{keywords}

\section{Introduction}

In a previous paper we reported on the discovery of an \ha\ burst in
the B emission line star HD 76534 (Oudmaijer \& Drew, 1997). During an
observing run in 1995, two days after a strong \ha\ emission line was
observed, the star was re-observed, and found to have only a
photospheric absorption line. Spectacularly,  two hours later, the
line was again in emission, still increasing its strength with respect
to the continuum.

Especially marked spectral variations have been detected in a few other Be
stars, notably $\mu$ Cen (Peters, 1986; Hanuschik \ea\ 1993, Rivinius
\ea \ 1998) and $\lambda$ Eri (Smith, Peters \& Grady 1991, Smith \ea
\ 1997a+b).  These two stars have long periods with relatively stable
\ha\ emission, and sometimes undergo an \ha\ outburst where the
emission grows to a maximum within days, displaying rapid variations
of the violet and red peaks of the \ha \ line.  The emission then
fades on longer timescales.

In contrast to both $\mu$ Cen and $\lambda$ Eri, the observed
timescale of the \ha\ outburst of HD 76534 was an order of magnitude
shorter, while the line profile did not show any V/R
variability. Instead, the line that was present only two hours after
the absorption was observed, was similar in profile and V/R ratio to
existing high resolution \ha \ spectra of the object, and did not
betray any signs of  on-going formation of recently ejected disk
material.

Two hypotheses were put forward by us to explain the rapid
variations. In analogy with the hypothesis for $\mu$ Cen, a sudden
burst of mass loss was first considered (see Hanuschik \ea \ 1993,
Rivinius \ea \ 1998), but based on the above arguments it was discarded
in favour of the idea that a stable rotating Keplerian disk was already
present around the star, but that a lack of ionizing photons from, for
example, the stellar photosphere failed to produce sufficient
ionizations and subsequent recombinations to push the \ha\ line into
emission. From simple considerations, it was found that a slight
change in ionizing flux can indeed ionize an existing stable neutral
disk, and result in detectable \ha \ emission.

It is not clear however where the change in ionizing radiation should
come from.  By analogy with the EUV variations of the $\beta$ Cep star
$\beta$ CMa (Cassinelli \ea \ 1996), it is possible that stellar
pulsations are responsible for this behaviour.  $\beta$ CMa shows
relatively large (30\%) variations in its Lyman continuum, which are
not as readily visible in the optical (Cassinelli \ea \ 1996).  A
similar effect could be happening in the case of HD 76534; at the
times when the Lyman continuum is at minimum, no \ha\ emission is
visible, while at maximum the line will develop.  A critical test of
the stellar pulsation hypothesis would be to monitor the star for
several hours up to several days to investigate whether any
periodicity would be present in the \ha \ emission of the object.

On the other hand, Smith \ea \ (1997b) reaches a similar conclusion to
explain, amongst other phenomena, the \ha \ variations in $\lambda$
Eri. A source of extra Lyman continuum photons could be responsible
for extra ionizations and recombinations in the circumstellar
material. Smith \ea \ (1997b) find that this can be explained
by the occurrence of heated slabs, possibly related to magnetic
activity, close to the stellar photosphere.  This activity does not
appear regularly, so no apparent periodicity in the \ha \ line,
certainly not on timescales of hours, would be expected.

Previous to Oudmaijer \& Drew (1997), observations of the \ha\ line of
HD 76534 were reported only twice, by Th\'e \ea\ (1985) and Praderie
\ea\ (1991).  Their spectra were taken one year apart, and showed
`indistinguishable' (cf.  Praderie \ea\ 1991) line profiles. We
measure an equivalent width (EW) from Th\'e \ea \ of --7 \ang .  The
exposure times used by these authors were 2 and 2.5 hours
respectively, so that any shorter term variations in either the line
profile or EW would have been washed out.

Since our previous paper, several new datasets on \hzz \ have been
published.  Recently, Corcoran \& Ray (1998), report a measurement of
the EW of the \ha \ of --14.3 \ang , on a spectrum that was obtained
in the last week of 1991 (their only Southern observing run) and
Reipurth, Pedrosa \& Lago (1996) show a spectrum, obtained in February
1993, with an EW of --10 \ang. These two papers do not mention the
exposure times. Oudmaijer \& Drew (1999) report on observations taken
in December 1995 and December 1996 and found an EW of --6 \ang \ and
--4 \ang \ respectively (we note a typographical error in their Table
2, the data for 30 December 1996 and December 1995 should be
interchanged).  It is clear that the \ha \ line of \hzz \ is variable
on long timescales.

The main aim of this paper is to present time-resolved observations of
the object at high resolution as a first step in determining whether
\ha\ is prone to any variations on timescales of minutes.  The
observations thus serve as a check whether the long integration times
by Praderie \ea\ (1991) and Th\'e \ea\ (1985) would indeed have washed
out any short-term variations.

This {\it Research Note} is organized as follows: In Sec. 2 we present
the time-resolved observations of the object obtained with the UCLES
spectrograph on the AAT.  In Sec. 3 we will search for variations in
the data and revisit the multi-epoch {\it V} band photometry obtained
by Hipparcos. We will conclude in Sec. 4.

\begin{small}
\begin{table}
\caption{Log of the observations 
\label{log}}
\begin{tabular}{rrrr}
\hline
UT start & t$_{exp}$ (s) & SNR & Zenith  \\
         &               &      & Distance (\degree)\\
\hline
09:03:19 &  60 &60 & 39 \\
09:05:35 &  60 &58 & 39 \\
09:07:51 &  60 &77 & 39 \\
09:10:07 &  60 &73 & 40 \\
09:12:24 &  60 &68 & 40 \\
09:14:50 & 150 &84 & 40 \\
09:18:36 & 150 &79 & 41 \\
09:22:23 & 150 &96 & 42 \\
09:26:09 & 150 &117& 43 \\
09:29:55 & 150 &114& 43 \\
09:34:05 & 240 &144& 44 \\
09:39:21 & 240 &138& 45 \\
09:44:37 & 240 &140& 46 \\
09:49:54 & 240 &119& 47 \\
09:55:10 & 240 &109& 48 \\
10:02:40 &  60 &56 & 49 \\
10:04:56 &  60 &56 & 50 \\
10:07:12 &  60 &65 & 50 \\
10:09:28 &  60 &61 & 51 \\
10:11:45 &  60 &71 & 51 \\
10:14:09 & 150 &86 & 52 \\
10:17:55 & 150 &100& 53 \\
10:21:41 & 150 &124& 53 \\
10:25:27 & 150 &89 & 54 \\
10:29:13 & 150 &102& 55 \\
10:33:04 & 240 &87 & 56 \\
10:38:21 & 240 &117& 56 \\
10:43:37 & 240 &93 & 57 \\
10:48:53 & 240 &77 & 58 \\
10:54:09 & 240 &84 & 59 \\
\hline
\end{tabular}
\ , \\
SNR measured in the range 6537 -- 6542 \ang \
\end{table}
\end{small}

\section{The Observations}

During the night of 7 June 1996 (UT), HD 76534 was observed in service
time, employing the UCLES spectrograph mounted on the 3.9 m
Anglo-Australian Telescope.  The observational set-up included the 31
grooves/mm echelle and a 1024$\times$1024 Tektronix CCD detector.  The
resulting spectrum contains 44 orders covering the spectral coverage
of 4650 to 7240 \AA, including both \ha\ and H$\beta$.  The observing
strategy was fairly straightforward, we aimed to obtain spectra every
few minutes, and employed different exposure times in order to
compromise between time-resolution and signal-to-noise.  Within a time
span of two hours, we obtained 30 spectra with signal-to-noise ratios
(SNR) in the \ha \ setting ranging from $\sim$ 60 in the shortest
(60s) exposures to $\sim$ 140 in the longest (240s) exposures. The
observing conditions were not ideal, so the total count-rates and SNR
are slightly variable.

Data reduction was performed in {\sc iraf} (Tody, 1993), and included
the procedures of bias-subtraction, flatfielding and wavelength
calibration.  An observation of a Th-Ar arc lamp was used to provide
the wavelength calibration of each spectrum.  The resulting spectral
resolution was determined to be $\sim$ 10 \km\ from Gaussian fits
through telluric absorption lines.

In total 30 spectra with different exposure times were taken. The
summed total spectrum, with a total exposure time of 4500s has a SNR
of $\sim$200 in the H$\beta$ order and $\sim$300 in the \ha \ order.
The log of the observations and the SNR in the \ha \ order  are provided
in Table~\ref{log}.

\section{Results}

\subsection{Short description of the spectrum}

Apart from the strong \ha\ emission (EW = --8 \ang ) and the filled-in
H$\beta$ absorption, no apparent emission lines are found. Strong He{\sc
i} lines at 4921, 5047, 5876 and 6678 \ang \ show absorption and allow for a
determination of the stellar  velocity,
which is found to be 14 $\pm$ 3 \km \ (heliocentric). This is
consistent with other determinations (see Oudmaijer \& Drew 1997).
Within the observational error-bars, we do not find evidence for
radial velocity variations.

\begin{figure}
\mbox{\epsfxsize=0.48\textwidth\epsfbox[30 160 515 640]{
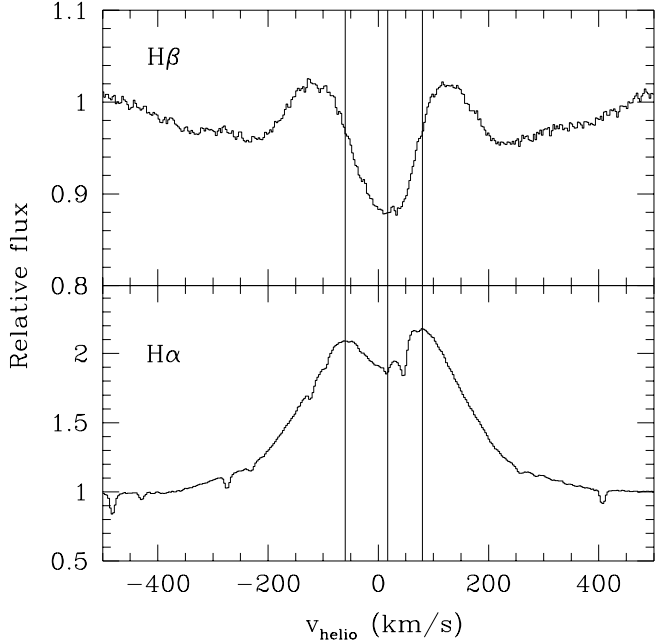}}
\caption{
The spectrum around \ha \ and H$\beta$.  The vertical lines are 
drawn through the \ha \ peaks  and the
systemic velocity. The larger peak separation in H$\beta$ 
suggests  that the hydrogen
recombination lines are  formed in a rotating  Keplerian type disk.
\label{halbet}
}
\end{figure}

The continuum corrected spectra around the \ha \ and H$\beta$ lines
are shown in Fig.~\ref{halbet}. \ha \ is a strong doubly-peaked
emission line, while the doubly-peaked H$\beta$ emission hardly
reaches above the local continuum.  The peak separation in the
H$\beta$ line is 242 $\pm$ 5 \km \ which is almost twice as large as
observed in the \ha \ line (143 $\pm$ 5 \km ).  The larger peak
separation in H$\beta$ is consistent with the fact that both lines are
formed in a rotating Keplerian type disk. If both lines, or \ha \ alone,
are optically thick, the \ha \ forming region is larger than the
H$\beta$ forming region, and will thus trace lower rotational
velocities.

\subsection{Variability on short time scales}

\begin{figure}
\mbox{\epsfxsize=0.48\textwidth\epsfbox[30 160 515 640]{
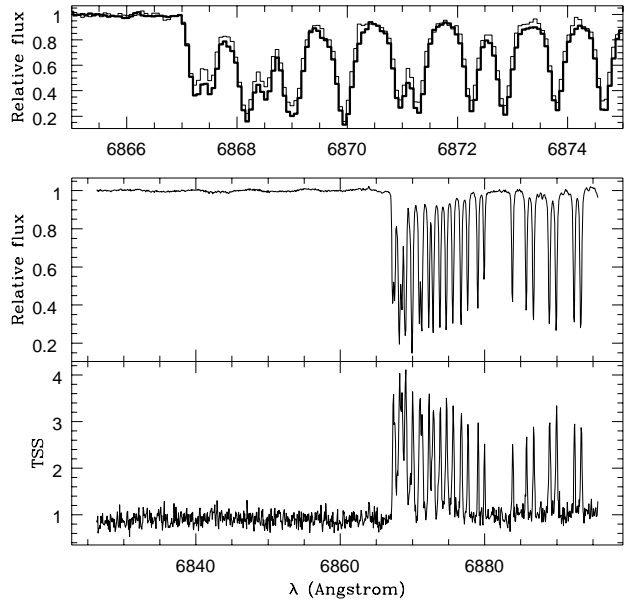}}
\caption{
The spectrum around the telluric lines at 6860 \ang . The upper panel
shows part of the spectrum, where the thick line represents the last
spectrum taken and the thin line the first spectrum, obtained almost
two hours earlier. It is clear that the telluric absorption has
become stronger. The middle panel shows the total, continuum
rectified, spectrum, and the lower panel presents the TSS. While the
continuum does not show any significant variations, the changes in the
telluric lines show up as strong peaks in the TSS.
\label{order05}
}
\end{figure}

\subsubsection{Method}

Visual inspection of the 30 individual spectra did not reveal any
obvious variability in the \ha\ line, so instead, we adopt the simple
statistical formalism presented by Henrichs \ea\,(1994), and already
exploited by Oudmaijer \& Bakker (1994) for a similar experiment for
the post-AGB star HD 56126.  This method was devised by Henrichs \ea \
to spot the regions of interest in their multi-epoch data on hot
stars. The variability can be expressed in a temporal variance
spectrum (TVS):

\begin{equation}
(TVS)_{\lambda} \approx \frac{1}{N-1} \sum_{i=1}^{N}\left(\frac{ F_{i}(\lambda) -
F_{av}(\lambda)}{\sigma_{i}(\lambda)}\right)^{2}
\end{equation}

where $N$ is the number of spectra, $F_{av}(\lambda)$ represents the
constructed average spectrum, $F_{i}(\lambda)$ the individual spectra, and
$\sigma_{i}(\lambda)$ =  $F_{i}(\lambda)$/(SNR) of each individual pixel of
the spectra.

Then, the so-called temporal sigma spectrum, TSS = $\sqrt{TVS}$), is
calculated.  This quantity represents approximately
($\sigma_{obs}/\sigma_{av}$). $\sigma_{obs}$ traces the standard
deviation of the variations of the individual spectra with respect to
the average spectrum, while $\sigma_{av}$ represents the standard
deviation of the average spectrum.  If no significant variations are
present in the spectra, the ratio of these two numbers,
$\sigma_{obs}/\sigma_{av}$, will be close to one, deviations are
directly represented in units of the noise level, that is to say a
peak `Temporal Sigma Spectrum' of three corresponds to a variability
at a 3$\sigma$ level.

During the extraction process, IRAF provides a SNR spectrum based on
the photon-statistics of the data. This is very convenient, as the SNR
changes strongly over a given order because the blaze of the
spectrograph results in lower count-rates and thus lower SNR at its
edges, while, of course, the countrates and SNR also change across
absorption and emission lines compared to the local continuum. As a
check, we measured the SNR in several wavelength intervals. The
IRAF-extracted SNR were scaled up by 40\% to bring the measured and
the IRAF SNR in agreement.  In the remainder of this exercise, we will
use these SNR spectra as input for Eq.~1.

The average spectrum was constructed by summing all individual
spectra.  After this, the spectra were continuum rectified as input
for Eq.~1. In the case of the \ha \ line, the wavelength region  used for the
fit were the blue and red continua beyond 13 \ang \ from the center of
the line. The $\sigma(\lambda)$ spectrum was computed by dividing the
individual continuum rectified spectra by their respective SNR
spectra.

Fig~\ref{order05} shows an example of the usefulness of the method. In
the middle panel the total spectrum in the order around the telluric
absorption bands at $\sim$ 6870 \ang \ are shown, the lower panel
shows the derived TSS. The continuum shows a trend from TSS $\sim$ 0.9
to $\sim$ 1.1, indicating a slight variation in the continuum of the
extracted spectrum.  The telluric lines vary however at the
2-4$\sigma$ level. This is due to the changing airmass during the
observations: as the airmass increases from 1.30 (zenith distance
39\degree ) to 2.0 (59\degree) the telluric absorption becomes
stronger. This is visible in the upper panel which shows an overplot
of the first and last spectrum taken.  The changes, which are only a
few percent of the continuum level, are real.

The fact that a slight variability is traced in the continuum
illustrates a very important caveat of the method. The TSS only
depends on photon-statistics, and is insensitive to any systematic
errors that may be present. In particular, a variable response curve
(`blaze') of the echelle, can show up as variability, while in reality
such variations are purely systematic rather than intrinsic. In the
case of Fig~\ref{order05} this is not so important, as the entire
order can be used for the continuum rectification, effectively
removing this effect.  In the case of \ha \ however, a large part of
the spectrum can not be used for the continuum rectification as, of
course, it is covered by the \ha \ line itself.

\begin{figure}
\mbox{\epsfxsize=0.48\textwidth\epsfbox[30 160 515 640]{
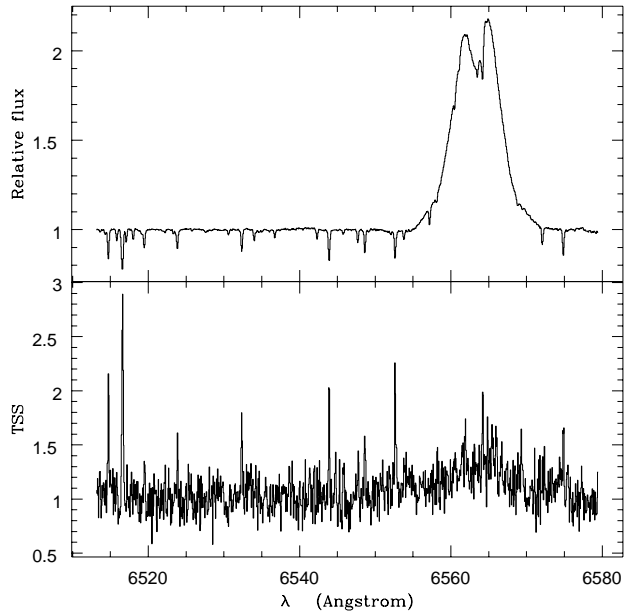}}
\caption{
Top panel: The continuum corrected total spectrum of the echelle order
covering \ha . Lower panel: TSS spectrum in the \ha \ order. The
telluric absorption lines again show up as healthy variable lines,
across the \ha \ line the variations are marginal.
\label{hal}
}
\end{figure}

\subsubsection{The \ha \ line}

Fig.~\ref{hal} shows the resulting TSS for the \ha \ order. The
telluric lines show variability at the 2.5$\sigma$ level. This is
smaller than the variability observed for telluric lines in
Fig.~\ref{order05}, but can be explained by the fact that these lines
and their changes are weaker.  It is nevertheless an important check
to note that the method also works in the \ha \ order.

\ha \ itself hardly shows any variability. In fact, the most
significant variability is due to the telluric absorption in the
central minimum of the line. The low $<$1.5$\sigma$ variability
observed across the line is statistically not significant.
Nevertheless, we investigated the possible cause of these marginal
variations. This was done bearing in mind the fact that a heavily
rebinned spectrum has a much larger SNR, and thus any variations 
should show up with greater significance.

Unfortunately, the echelle orders' wavelength ranges are rather
limited (about 67 \ang \ or 3000 \km \ in the \ha \ order) compared to
the extent of the line itself (full-width at zero intensity $\sim$
1000 \km ). It is thus possible that systematic variations in the
continuum interpolated underneath \ha \ may be mis-interpreted as
revealing intrinsic variations in the line.  Indeed, by dividing all
individual spectra around \ha \ by the same (rescaled) continuum fit,
it became clear that the curvature of the spectra varies in time on a
level less than a few \%, having biggest impact on the red end of the
spectrum.  This is probably related to a well-known varying blaze due
to the de-rotator optics in UCLES.

In order to check whether the line may be intrinsically variable, we
performed some tests. The main reasoning behind these
tests is that if the response curve of the echelle is variable in
time, the adjacent (line-free) orders should show a similar
variability. We therefore investigated the two orders next to the \ha
\ order in the echellogram, and continuum rectified these using the
same pixel range as the \ha \ order, i.e. not using the $\sim$ 25 \ang
\ around the center of \ha , and looked for evidence of variability.

We measured the EW of a fiducial line over the same pixel-range in
these orders (corresponding to 26 \ang ) as \ha . The measured EW in
both orders is close to 0 \ang , but has a scatter of 0.12 \ang . This
is to be compared with the scatter in the EW of the \ha \ line of 0.21
\ang . Based on the variations of the EW of the fiducial lines in the
continuum of the adjacent echelle orders and the mean height of \ha \
line over the measured interval, we would expect a scatter of 0.16
\ang \ in the EW of \ha .  The scatter of the EW of \ha \ is slightly
larger than this, corresponding to variability at a 1.3 $\sigma$ level.

Having established that the total \ha \ EW is hardly variable, the
question is whether this is because the total line is not variable at
all, or whether the line-profile changes in such a way that the total
line-flux is nearly constant. Checks on the individual spectra show
that the small changes in the line-profile are in phase with each
other, and more importantly, in phase with changes in the red
continuum flux. This indicates that the line-profile as such does not
vary, while it traces the changes in the continuum level. Hence the
insignificant variability in the EW is not due to changes in the
line-profile.

Based on the facts that the EW of the \ha \ line changes at a similar
amplitude as the EW in the same pixel-range of the adjacent orders,
and that the `changes' traced by the TSS spectrum are in phase with
the red continuum, we conclude that during the two hours of these
observations, no significant variability was present in the \ha \ line
of \hzz .

\begin{figure}
\mbox{\epsfxsize=0.48\textwidth\epsfbox[30 160 515 640]{
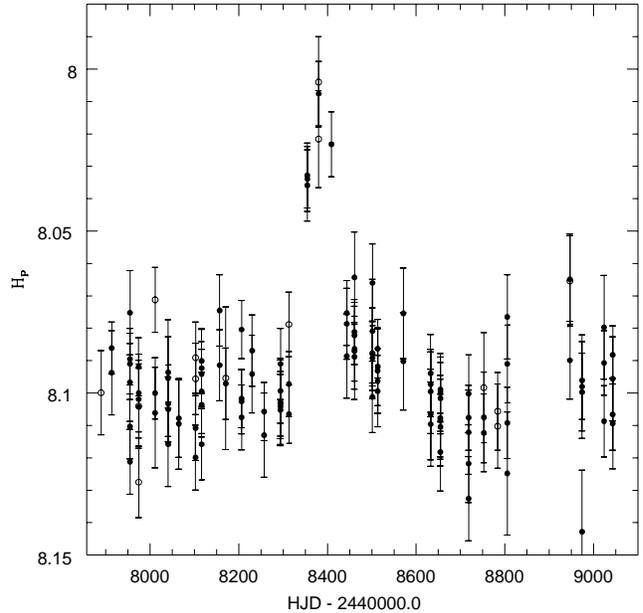}}
\caption{
Hipparcos photometry of HD 76534. The open circles are the data with
HT4 $>$ 0, indicating, mostly in this case, that only one of the two
consortia obtained this value.
\label{hip}
}
\end{figure}

\subsection{Hipparcos photometry re-visited}

The Hipparcos satellite observed HD 76534 125 times between 1989 and
1993 photometrically in a passband similar to the {\it V} band (ESA
1997).  These observations were reported on in the paper on Herbig
Ae/Be stars by van den Ancker, de Winter \& Tjin A Djie (1998).  These
authors mentioned that the star is probably photometrically variable,
which they based on the fact that the variance of the individual
photometric points is larger than the observational errors. No
lightcurves were provided however.

Here we look at the data provided by the Hipparcos catalogue into more
detail.  The photometry is plotted as function of Julian Date in
Fig.~\ref{hip}.  In the first year of the mission, \hzz \ was constant
within the errorbars until the object brightened by about 0.1
magnitude, reaching a maximum around May 1991. The period of
brightening and fading lasted about 100 days. Afterwards the object
`flickers' around a mean value close to what was measured before the
maximum.

Could this rise in brightness be associated with the spectral
behaviour of the object?  Mennickent \ea \ (1998) published 11 year
long photometric monitoring of $\lambda$ Eri, and found several
similar changes in the Str\"omgren photometry of the object. A period
search revealed that rises in brightness of $\sim$0.1 mag. occurred
with a period of 486 days, while the rising and fading of the object
lasted about 100 days. From the colour changes, they found that the
brightness maxima correspond to slight increases in effective
temperature of the star.  Although the overlap between spectroscopic
and photometric data is not very complete,  Mennickent \ea \ find a
rough correlation between the jumps in brightness and periods of \ha \
emission in the star.

\section{Concluding remarks}

In this paper we have investigated whether the \ha \ flaring star \hzz
\ is variable on short timescales of order minutes and hours.  We find
no evidence for statistically significant variations.  One of our
original suggestions to explain the \ha \ outburst in 1995 was to
invoke a pulsation-type behaviour of the star. If this would have been
the case, we might have seen at least some variability on short time
scales, but these are not seen.

Including the current data, \ha \ measurements for this object have
now been reported 9 times in the astronomical literature (see the
Introduction), and all but one show the line in emission. Considering
this, one would conclude that the observed collapse and the subsequent
rapid recovery of the \ha \ emission is only a sporadic event. If we
were to associate the brightening of the object in the Hipparcos
photometric data to a strong emission variability, it may be that the
relevant timescale is the hundreds of days between events. A
monitoring programme sampling a range of timescales is needed to clear
this up.

In terms of variability, HD 76534 can be regarded as belonging to the
same class as $\mu$ Cen and $\lambda$ Eri. But  there are some
striking differences. In the first place, the \ha \ variations in 1995
were the strongest and fastest ones observed to date in a Be
star. Secondly, whereas $\mu$ Cen and $\lambda$ Eri are most often
observed in a quiescent state, interrupted by bursts of emission that
gradually declines, HD 76534 is most often observed with \ha \
emission, interrupted by absorption, suggesting a different process
governing the \ha \ behaviour. Indeed, the \ha \ line that was
observed only two hours after it had been in absorption showed a
smooth doubly-peaked emission profile, similar to what has been
observed now, suggesting that a neutral disk was present before the
source of ionizing photons increased in strength contrary to the
common idea that a disk has been built up in the case of $\mu$ Cen and
$\lambda$ Eri.

\paragraph*{\it Acknowledgments:}

We thank Myron Smith for interesting and inspiring discussions.  We
thank the staff at the Anglo-Australian Telescope for their expert
advice and support.  RDO is funded by the Particle Physics and
Astronomy Research Council of the United Kingdom. The data analysis
facilities are provided by the Starlink Project, which is run by CCLRC
on behalf of PPARC.  This research is partly based on results from the
ESA Hipparcos astrometric satellite.

\end{document}